\documentclass[prl,twocolumn, preprintnumbers]{revtex4}
\usepackage{amssymb,amsmath}
\renewcommand{\ggg}{\ensuremath{\gamma\rightarrow 3\gamma}}
\newcommand{\less}{<}

\begin{document}
\begin {center}
\preprint{
UCLA/05/TEP/10
}\end{center}
\vspace*{2mm}

\title{
 On photon splitting in theories with\\
Lorentz invariance violation.
}

\author{Graciela Gelmini$^1$}
\author{Shmuel Nussinov$^2$}
\author{Carlos E. Yaguna$^1$}
\affiliation{$^1$ Department of Physics and Astronomy, University
of California, Los Angeles, CA 90095-1547,USA}

\affiliation{$^2$ Sackler Faculty of Science, Tel Aviv University,
 Tel Aviv, Israel}

\begin{abstract}
In a model with Lorentz invariance violation implemented through
 modified dispersion relations,  we estimate the rate 
for the decay process $\gamma\rightarrow 3\gamma$  and find that it
provides a relevant bound on Lorentz invariance violation.

\vspace{.3cm}


\end{abstract}

\maketitle
Lorentz invariance is not necessarily an exact symmetry of nature. 
Though all the experimental results show that the laws of physics 
are Lorentz invariant, several approaches to quantum gravity, for 
example, suggest that at very high energies Lorentz invariance 
might be broken. Surprisingly, such violations, even if associated
with Planck scale physics, may be constrained by present observations.

To discuss Lorentz invariance violation (LIV) we adopt the kinematic
framework based on modified dispersion relations. That is, we alter
 the  relation between the energy and the momentum of a given particle
 but keep intact the dynamics and the energy-momentum conservation law.
 Such modifications are indeed suggested by certain approaches to
 Quantum Gravity~\cite{Vucetich:2005ra}.

Modified dispersion relations manifest in different 
ways~\cite{Jacobson:2002hd}. They may affect the propagation 
of photons by inducing dispersion and birefringence over long 
travel times. They could  modify the threshold for known processes,
 as in $\gamma\gamma\rightarrow e^+e^-$ and the GZK 
reaction~\cite{Coleman:1998ti}. They might also induce new processes 
not allowed in QED such as photon decay ($\gamma\rightarrow e^+e^-$), 
photon splitting
($\gamma\rightarrow \gamma\gamma, \gamma\gamma\gamma, ... $), and vacuum Cerenkov
radiation ($e\rightarrow e\gamma$). The non-observation of such
anomalous effects, then, constrains the assumed dispersion relations.

We consider, following Ref.~\cite{Jacobson:2002hd}, the modified dispersion 
relation for the photon
\begin{equation}
   E_\gamma^2=p_\gamma^2 + \xi \frac{p_\gamma^3}{M}
\label{dispersion}
 \end{equation}
where $E_\gamma$ and $p_\gamma$ are the energy and momentum of the photon,
the scale  $M =10^{19}$~GeV is an energy scale close to the Planck mass, 
and $\xi$ is a Lorentz invariance violating parameter.
 This dispersion relation preserves rotation invariance but  not boost
 invariance. Thus it can hold in only one reference frame, usually
identified with the rest frame of the  cosmic microwave background.

Within an effective field theory (EFT) framework, the photon 
satisfies the polarization dependent dispersion relation 
$E_{\pm}^2=p^2\pm \xi (p^3/M)$, where the subscripts $\pm$ refer to the right 
and left circular polarizations of the photon~\cite{Kostelecky:2002hh}. 
It may be, however, that EFT does not describe the leading effects of
 Lorentz invariance violation. In any case, EFT is a dynamical assumption
 that goes beyond the kinematic framework we are following. We will, 
therefore, neglect polarization dependence in the dispersion relation 
and assume that all photons satisfy Eq.~(\ref{dispersion}) with $\xi$ positive.

In standard QED the photon splitting process $\ggg$ does not occur because,
due to energy and momentum conservation, the three momenta must all be parallel and
the amplitude vanishes in this configuration. However, such decay is allowed in the
presence of the modified dispersion relation in Eq.~(\ref{dispersion}) for
$\xi>0$. Since in this case the process $\ggg$ has no threshold, 
its effectiveness in constraining
$\xi$ depends solely on the rate. So far, only a crude guess of
this rate exist in the literature. In this letter, we provide an estimate
of the decay rate $\Gamma(\gamma\rightarrow 3\gamma)$ 
and show that the observation of high energy photons from astrophysical 
sources provides an upper bound on $\xi$ stronger than previously believed.

To compute the rate of the decay $\gamma\rightarrow 3\gamma$ 
we write the modified dispersion relation in terms of an ``effective photon mass",
\begin{equation}
m_\gamma^2=\xi\frac{p_\gamma^3}{M}\,,
\label{mass}
\end{equation}
so that the photon satisfies $E_\gamma^2=p_\gamma^2+m_\gamma^2$.

Notice that all the scalar products of photon momenta in the process $\ggg$,
as well as all the scalar products of momenta and polarization vectors,
vanish for $m_\gamma=0$. The presence of a small $m_\gamma$ ( $\not= 0$)
changes the kinematics of the process, allowing the final photon momenta to have
small components perpendicular to the initial momentum and proportional to $m_\gamma^2/ E_\gamma$. In addition, each photon acquires a longitudinal polarization vector $\epsilon^\mu\sim p_\gamma^\mu/m_\gamma$.
As a result, all the non-vanishing scalar products of photon momenta
as well as all the  scalar products of momenta and polarization vectors
are of the order of the initial effective photon mass. Thus, apart from the  dependence
of the decay rate on $1/ E_\gamma$, which remains explicitly in the final result, the
important energy scale for the rate is given by the ``effective mass" of the initial photon (the
``effective photon masses" of the final photons  are smaller and typically of the same order of magnitude of the initial photons mass, due to their dependence on  the photon momenta).

We believe that this argument shows clearly that the relevant energy scale for photon decay is not
the energy $E_\gamma$ of the initial photon, but its ``effective mass". And due to the
large suppression $M^{-1}$ of this effective mass  in Eq.~(\ref{mass}),
even energetic photons
may have a small effective mass
(say $m_\gamma\sim 10^{-2}\xi^{1/2} \mbox{MeV}$ for $E\simeq \mbox{TeV}$).

 To proceed with our argument,  let us define the ``effective initial photon gamma factor" $\gamma = m_\gamma/ E_\gamma$, multiply and divide the standard equation defining the differential decay rate by $m_\gamma$ and extract a factor of
inverse $\gamma$ from it. The remaining equation is identical to the decay rate computed in the reference frame where the energy of the initial photon is $m_\gamma$ (which is the ``rest frame" of our ``massive photon"), if we compute the amplitude
from Lorentz invariant terms (there may be other non-Lorentz invariant terms but certainly the contribution of the Lorentz invariant terms should be there). In fact, if the squared amplitude
$|M_{\beta\alpha}|^2$ term is invariant (here $\alpha$ stands for the initial state and $\beta$ for the final state) we can change frames in the usual manner to compute the
decay rate, since the other factors are $\delta^{(4)}(p_\beta-p_\alpha)$, still valid since we are assuming energy and momentum conservation, and the invariant volume of the final particles, which would not be affected with respect to the standard case if we consider that only the initial photon has an ``effective non-zero mass"
(in any event the order of magnitude of the decay rate would not be changed by taking effective photon masses as defined above also for the final photons or only for the initial photon).

We think that these arguments justify to estimate the decay rate of this ``massive photon" in its
rest frame and boost it -with the usual Lorentz factor- to the laboratory
frame (which practically coincides with the rest frame of the CMB)
to obtain the desired rate. This calculation accounts for the contributions to the decay rate of the Lorentz invariant terms in the Lagrangian (those present in the  usual QED) although there may be
other LIV terms in the Lagrangian~\cite{Jacobson:2005bg}, in particular those terms of dimension five
which give origin to the modified dispersion relations we consider. The contribution of such  terms, however, seems to be smaller than the one we computed here~\cite{Jacobson:2002hd, Jacobson:2005bg}.

The ``effective photon mass" will need to be smaller than the electron mass (as shown below, $m_\gamma^2 < 4 m_e^2$) for the photon splitting
to dominate the decay rate over the process $\gamma\rightarrow e^+ e^-$. In fact we will find that
the ``effective photon mass" is
 much smaller than the electron mass. Thus the electron mass should not be neglected in the dimensional
arguments given in Ref.~\cite{Jacobson:2002hd}. In this reference it is inferred that the decay rate must
depend of a factor $E^5 \xi^4 M^{-4}$ at photon energies $E$ well above the electron mass. However, as we have shown,  it is the effective mass of the photons which provide the relevant energy scale, and not the initial photon energy. We find here that the rate contains an additional factor
of $(E_\gamma/m_e)^8 \xi M^{-1}$, which comes from the inverse gamma factor $(m_\gamma/ E_\gamma)$ times
a factor $(m_\gamma/m_e)^8 $ due to the Euler-Heisenberg Lagrangian we use
to describe the interactions among photons. This  Lagrangian provides the effective 4-$\gamma$
coupling given by QED at energy scales smaller than the electron mass. Note that we are not modifying QED to actually include a photon mass in the Lagrangian. We use the modified kinematics induced by
an ``effective photon mass", but do not modify the dynamics.


In order to gain confidence in the  procedure we propose, we will start by
applying it to two processes for which rate calculations have been published,
i.e.  $\gamma\rightarrow e^+ e^-$ and $p \to p \gamma$~\cite{Coleman:1997xq}.
Let us first consider   $\gamma\rightarrow e^+ e^-$.
 In the center of mass frame, the decay rate of a ``massive photon"
 into an electron-positron pair is simply
$\Gamma_{CM}\simeq (\alpha/2) m_\gamma$. In the laboratory
frame the rate  gets an additional inverse gamma factor,
 $\gamma^{-1}=m_\gamma/E_\gamma$,
\begin{equation}
\Gamma_{lab}=\frac{\alpha}{2} \frac{m_\gamma^2}{E_\gamma}=
\frac{\alpha}{2}  \frac{\xi E_\gamma^2}{M}\,.
\end{equation} 
This result coincides with that obtained in Ref.~\cite{Coleman:1997xq},
which is  $\Gamma\simeq ({\alpha}/{2}) (c^2-1) E$, 
for a photon
with four momentum $(E,E/c)$, if we replace the effective photon mass in
Eq.~(3) by its value in the model of  Ref.~\cite{Coleman:1997xq}, namely 
 $m_\gamma^2= E^2(c^2-1)$. The energy threshold for this
 decay is obviously given by the condition $m_\gamma^2=4 m_e^2$, which
 is equivalent to the condition $E^2=4 m_e^2/(c^2-1)$ found in 
Ref.~\cite{Coleman:1997xq}. Thus, in terms of the ``effective
 photon mass" the expression for the energy threshold and the decay
rate are particularly simple. In fact,
 in the footnote 7 of Ref.~\cite{Coleman:1998ti}, the procedure we follow is mentioned
as providing the right result.

The rate for the $p\rightarrow p\gamma$ decay
of Ref~\cite{Coleman:1997xq}  can also be obtained from
kinematic considerations similar to those used above for $\gamma\rightarrow e^+ e^-$.
Let us use
the modified dispersion relation for the proton also proposed in Ref.~\cite{Jacobson:2002hd},
\begin{equation}
E_p^2=p_p^2+m^2+\eta \frac{p_p^3}{M}~.
\label{proton}
\end{equation}
Since the effective mass of the proton depends on its momentum, the mass
 of the parent proton, $m_P$, is larger than that of the daughter $p$, $m_D$,
and the decay $p\rightarrow p \gamma$ is kinematically allowed. In the
rest frame of the  parent proton the rate is
$\Gamma_{CM} \simeq \alpha m_P (1- m_D^2/m_P^2)$.
In the laboratory frame the rate  gets an additional inverse gamma factor,
 $\gamma^{-1}=m_P/E_p$, thus
\begin{equation}
\Gamma_{p\rightarrow p \gamma}\simeq \alpha
\frac{m_P^2}{E_p}(1- m_D^2/m_P^2)\simeq \alpha \frac{\eta E_p^2}{M}~.
\end{equation}
The result quoted in Ref.~\cite{Coleman:1997xq} is for the rate
of energy loss,   ${dE}/{dx}\simeq -\alpha E_p^2 (c_p- 1)$ (see the
notation of Ref~\cite{Coleman:1998ti} in which we have taken the maximal
possible speed of a  photon  $c_\gamma=1$ and the maximal speed of the
proton to be $c_p >1$).
Here $c_p=\partial E/\partial p$, and from Ref.~\ref{proton}
$c_p\simeq 1+\eta E_p/M$. Since $dE/dx\simeq - \Gamma E_p$,
the result of Ref.~\cite{Coleman:1997xq} is the same we obtain.

Let us now concentrate on $\ggg$. At energy scales (given by the ``effective photon mass") smaller than the electron mass, the interactions among photons are described
by the Euler-Heisenberg Lagrangian,
\begin{equation}
\mathcal{L}_{E\mbox{-}H}=\frac{2 \alpha^2}{45 m_e^4}
\left[\left(\frac 12 F_{\mu\nu}F^{\mu\nu}\right)^2 
+7 \left(\frac 18 \epsilon^{\mu\nu\rho\sigma} 
F_{\mu\nu}F_{\rho\sigma}\right)^2\right]~.
\label{euler}
\end{equation}
Using this  Lagrangian we  estimate $\Gamma(\gamma\rightarrow 3\gamma)$ 
in the laboratory frame (i.e. the CMB rest frame)  as

\begin{eqnarray}
\Gamma(\gamma\rightarrow 3\gamma)&
=&\left(\frac{2\alpha^2}{45}\right)^2 \frac{1}{3\mbox{!}\,
 2^{11} \pi^9}\frac{m_\gamma^{9}}{m_e^8}\frac{m_\gamma}{E_\gamma}\times f\\
&=& 1.5\times 10^{-20} \frac{\xi^5 E_\gamma^{14}}{M^5 m_e^8}\times f~,
\end{eqnarray}
where $f$ is the integral over momenta with some large factors
taken out, i.e.
\begin{equation}
f=\frac{4 \pi^4}{m_\gamma^9}\int 
\frac{d^3k_1\,d^3k_2\,d^3k_3}{E_1E_2E_3}
\delta^4(p_\gamma-k_1-k_2-k_3)\left|\mathcal{M}\right|^2,
\end{equation}
 and  $\mathcal{M}$ is the invariant matrix element 
obtained from Eq.(\ref{euler}) by omitting the global factor 
${2 \alpha^2}/({45 m_e^4})$.

Defined in this manner the factor $f$ is of order one. 
Indeed, a similar (but not identical)
 integration was performed numerically in Ref.~\cite{Adam:2002rg} and
 $f$ turned out to be 0.2. In our formulas we will keep
 explicitly the dependence on $f$ and we will see that our final 
bound depends very weakly on $f$ (in fact as $f^{-1/5}$, thus 
only an $f$  of order either 10$^5$ or 10$^{-5}$ would change the order of magnitude of our calculation).

Thus the photon lifetime is 
\begin{equation}
\tau(\ggg)=0.025 \xi^{-5}\,f^{-1}\,
\left(\frac{50 \mbox{TeV}}{E_\gamma}\right)^{14} {\rm sec}.
\label{lifetime}
\end{equation}

For a given $\xi$ and a particular photon time of flight 
$t$, the condition $\tau \sim t$ defines a critical value 
for the photon energy, $E_{c}$. Due to the strong dependence
 of the lifetime on the energy, photons with energies 
above $E_{c}$ would quickly cascade down  before reaching
  Earth from a distance $ct$ and therefore would not be observed. Thus, 
 photons reaching Earth from a distance $ct$ should satisfy
 $E_\gamma < E_{c}$ or equivalently $\tau  \geq t$.
 Since photons with $E_\gamma\approx 50 $TeV coming
 from the Crab nebula -$10^{13}$ seconds away- have
 been detected~\cite{Tanimori:1997cq}, we get the constraint
\begin{equation}
\xi\leq 1.2 \times 10^{-3} f^{-1/5} 
\left(\frac{50 \mbox{TeV}}{E_\gamma}\right)^{2.8}\,.
\label{constraint}
\end{equation} 

This bound is considerably more restrictive than the
bound $\xi\leq 10^{4}$ previously obtained~\cite{Jacobson:2002hd}
using the same decay mode, and it is comparable to the strongest constraint
on $\xi$, i.e. $|\xi|\lesssim 10^{-4}$~\cite{Gleiser:2001rm}, coming from
vacuum birefringence (different speeds for different photon
 polarizations). This latter bound, however, was derived within the EFT framework,
in which the Lorentz invariance violating parameters for left
 and right circular polarized photons have opposite signs.
Notice that the factor $f$ that we do not compute in this paper would
need to be $\sim 10^{-35}$ to bring the bound we obtain here to coincide with
the bound mention in Ref.~\cite{Jacobson:2002hd}, but since we have extracted from the
integral defining $f$ all large dimensional quantities, we believe that the remaining
integral cannot either very large or very small, but of order one.

Notice that as soon as the decay $\gamma \rightarrow e^+ e^-$ is
kinematically allowed, i.e. for  $m_\gamma^2 > 4m_e^2$, this decay mode,
that happens at tree level, dominates
 over the one-loop suppressed $\gamma\rightarrow 3\gamma$
We need to make sure that
  $m_\gamma^2\less 4m_e^2$, because then
  is the decay $\gamma\rightarrow 3\gamma$ important.
The condition $m_\gamma^2<4 m_e^2$, which
guarantees the validity of our approach, translates into
 a restrictive bound on $\xi$ less restrictive than the bound found above
\begin{equation}
\xi \leq 10^{-1} \left( \frac{50\,\mbox{TeV}}{E_\gamma}\right)^3.
\label{ee}
\end{equation}
Thus the required condition $m_\gamma^2\less 4m_e^2$ is automatically fulfilled.
This last condition insures also that the high energy photons coming
 from the Crab nebula
did not decay into $e^+e^-$ before they
reached Earth.

In conclusion, we have estimated the decay rate of the process
 $\ggg$ for photons fulfilling the Lorentz invariance violating
 dispersion relation $E^2=p^2+\xi p^3/M$ and found that the observation
 of high energy photons from the Crab nebula  sets
an important constraint on the  Lorentz invariance violation parameter
 $\xi$, much stronger than previously claimed.


\end{document}